\documentclass[twocolumn]{article}

\usepackage{graphicx,epsfig,subfigure}
\usepackage{amsmath,amssymb,cite}

\title{Fluctuation theorems in presence of information gain and feedback}
\date{}
\author{Sourabh Lahiri$^1$, Shubhashis Rana$^2$ and A. M. Jayannavar$^3$}

\begin{document}

%\maketitle{}

\newcommand{\nwc}{\newcommand}
\nwc{\la}{\langle}
\nwc{\ra}{\rangle}
\nwc{\lw}{\linewidth}
\nwc{\nn}{\nonumber}
\nwc{\Ra}{\Rightarrow}
\nwc{\dg}{\dagger}

\nwc{\Tr}[1]{\underset{#1}{\mbox{\large Tr}}~}
\nwc{\pd}[2]{\frac{\partial #1}{\partial #2}}
\nwc{\ppd}[2]{\frac{\partial^2 #1}{\partial #2^2}}

\nwc{\zprl}[3]{Phys. Rev. Lett. ~{\bf #1},~#2~(#3)}
\nwc{\zpre}[3]{Phys. Rev. E ~{\bf #1},~#2~(#3)}
\nwc{\zpra}[3]{Phys. Rev. A ~{\bf #1},~#2~(#3)}
\nwc{\zjsm}[3]{J. Stat. Mech. ~{\bf #1},~#2~(#3)}
\nwc{\zepjb}[3]{Eur. Phys. J. B ~{\bf #1},~#2~(#3)}
\nwc{\zrmp}[3]{Rev. Mod. Phys. ~{\bf #1},~#2~(#3)}
\nwc{\zepl}[3]{Europhys. Lett. ~{\bf #1},~#2~(#3)}
\nwc{\zjsp}[3]{J. Stat. Phys. ~{\bf #1},~#2~(#3)}
\nwc{\zptps}[3]{Prog. Theor. Phys. Suppl. ~{\bf #1},~#2~(#3)}
\nwc{\zpt}[3]{Physics Today ~{\bf #1},~#2~(#3)}
\nwc{\zap}[3]{Adv. Phys. ~{\bf #1},~#2~(#3)}
\nwc{\zjpcm}[3]{J. Phys. Condens. Matter ~{\bf #1},~#2~(#3)}
\nwc{\zjpa}[3]{J. Phys. A ~{\bf #1},~#2~(#3)}
\nwc{\zpjp}[3]{Pram. J. Phys. ~{\bf #1},~#2~(#3)}
\nwc{\zpa}[3]{Physica A ~{\bf #1},~#2~(#3)}

\twocolumn[
\begin{@twocolumnfalse}
  \maketitle{}
  \begin{center}
    Institute of Physics, Bhubaneswar - 751005, Sachivalaya Marg, India.
  \end{center}
  \begin{abstract}
In this study, we rederive the fluctuation theorems in presence of feedback,  by assuming the  known Jarzynski equality and detailed fluctuation theorems.  We first reproduce the already known work theorems for a classical system, and then extend the treatment to the other classical  theorems. For deriving the extended quantum fluctuation theorems, we have considered open systems. No assumption is made on the nature of environment and  the strength of system-bath coupling. However, it is assumed that the measurement process involves classical errors.

\vspace{0.5cm}
  \end{abstract}

\end{@twocolumnfalse}
\vspace{1cm}
]

\let\thefootnote\relax\footnotetext{$^1$lahiri@iopb.res.in}
\let\thefootnote\relax\footnotetext{$^2$shubho@iopb.res.in}
\let\thefootnote\relax\footnotetext{$^3$jayan@iopb.res.in}

\section{Introduction} 

In the last couple of decades,  active  research is being pursued in  the field of nonequilibrium statistical mechanics.  Until recently, the systems that are far from equilibrium had always eluded exact analytical treatments, in contrast to the well-established theory in the linear response regime.  Several new results have been discovered for systems that are far from equilibrium.  These relations are generically grouped under the heading {\it fluctuation theorems} \cite{eva93,eva94,jar97,cro98,cro99}. One of the major breakthroughs has been the Jarzynski Equality\cite{jar97}, which states the following. Suppose that the system, consisting of a Brownian particle, is initially prepared in canonical equilibrium with a heat bath at temperature $T$. Now we apply an external  protocol $\lambda(t)$ that drives the system away from equilibrium. The following equality  provided by Jarzynski is valid for this system \cite{jar97}:
\begin{equation}
\la e^{-\beta W}\ra = e^{-\beta\Delta F}.
\label{JE}
\end{equation}
Here, $W$ is the thermodynamic work defined by $W\equiv \int_0^\tau (\partial H_\lambda/\partial \lambda)\dot\lambda ~dt$. $H_\lambda$ is the Hamiltonian with externally controlled time-dependent protocol $\lambda(t)$, and $\tau$ is the time for which the system is driven.  $\Delta F$ is the difference between the equilibrium free energies of the system at the parameter values $\lambda(0)$ and $\lambda(\tau)$, which equals the work done in a reversible process. A direct outcome of the above equality is the second law of thermodynamics, which states that the average work done on a system cannot be less than the change in free energy: $\la W\ra\ge \Delta F.$  A further generalization of the above result was provided by Crooks \cite{cro98}. The Crooks' work theorem says that  the ratio of the  probability of the performing work $W$ on the system, $P_f(W)$, along the forward process and that of performing work $-W$ (i.e., extracting work from the system) along the time-reversed process, $P_r(-W)$, is exponential in the forward work, provided the initial state of either process is at thermal equilibrium:
\begin{equation}
\frac{P_f(W)}{P_r(-W)} = e^{\beta(W-\Delta F)}.
\end{equation}
Here, the initial probability density function (pdf) of the time-reversed process is the thermal/Boltzmann pdf corresponding to the final protocol value $\lambda(\tau)$ of the forward process. It is crucial to note that the fluctuation theorems are in complete conformity with the second law, since the {\it  average} work always exceeds the free energy, although for individual trajectories this condition may not be meted out \cite{lah11}.

The above theorems are valid for what are known as {\it open-loop} feedback, i.e., when the protocol function for the entire process is predetermined. In contrast, in a {\it closed-loop} feedback, the system state is measured along the forward trajectory, and the protocol is changed depending on the outcomes of these measurements. For such feedback-controlled systems, the fluctuation theorems need modifications so as to account for the information gained through measurement. Sagawa and Ueda have derived these extended relations for both the classical \cite{sag10,sag11} and  the quantum \cite{sag08} cases. In the original papers, a single measurement (at some instant $t=t_m$) was considered. Subsequently, in a detailed review \cite{sag11}, the authors have derived the relations in the classical case, when multiple measurements are being performed. 

In this paper we rederive the results for the classical systems, assuming the known fluctuation theorems in their integral as well as detailed form. The same treatment goes through for deriving the generalized Hatano-Sasa identity, which provides equalities for a driven system from one steady state to another. We also extend the same treatment to the quantum case, and show that no matter how many intermediate projective measurements and subsequent feedbacks are performed, the extended Tasaki-Crooks fluctuation theorem remains unaffected. The  efficacy parameters for classical and quantum systems are also obtained. We believe that our treatment is simple, as it assumes the already known Jarzynski Equality and the other fluctuation theorems in the absence of feedback.

\section{The System} 
\begin{figure}[h]
  \centering
  \includegraphics[width=7cm]{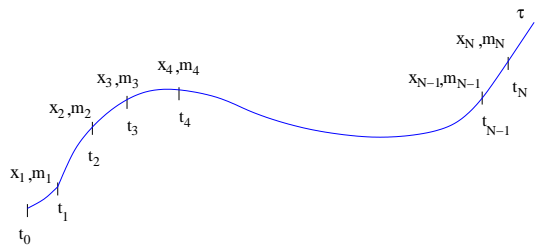}
  \caption{A typical phase space trajectory $x(t)$ vs $t$. The actual and the measured states at time $t_k$ are $x_k$ and $m_k$, respectively.}
\label{traj}
\end{figure}

We have a Brownian particle that is initially prepared in canonical equilibrium with a heat bath at temperature $T$. Now, we apply an external drive $\lambda_0(t)$ from time $t_0=0$ up to $t=t_1$. At $t_1$, we measure the state of the system and find it to be $m_1$ (see figure \ref{traj}). Then, we modify our protocol from $\lambda_0(t)$ to $\lambda_{m_1}(t)$ and evolve the system up to time $t_2$, where we perform a second measurement with outcome $m_2$. Subsequently the protocol is changed to $\lambda_{m_2}(t)$, and so on up to the final form of the protocol $\lambda_{m_{N}}(t)$, which ends at $t=\tau$ (total time of observation). However, the time instants at which such measurements are taken need not be equispaced. We assume that there can be a measurement error with probability $p(m_k|x_k)$, where $m_k$ is the measurement outcome at time $t_k$, when the system's actual state is $x_k$. Obviously, the value of $\Delta F$ will be different for different protocols $\lambda_{m_k}(t)$. 

\section{Extended Jarzynski Equality}
\label{ext_JE}
For a {\it given} set of observed values $\{m_k\}\equiv \{m_1, m_2, \cdots, m_N\}$, we have a fixed  protocol $\{\lambda_{m_k}(t)\}$ which depends on all the measured values $\{m_k\}$, as explained above. For such a given protocol trajectory, the Jarzynski Equality must be satisfied. Eq. (\ref{JE}) can be rewritten as
\begin{align}
&\int ~\mathcal{D}[x(t)]~p_{eq}(x_0)~P_{\{\lambda_{m_k}\}}[x(t)|x_0] \nn\\
&\times \exp\left\{-\beta W[x(t), \{m_k\}]+\beta\Delta F(\lambda_{m_N}(\tau))\right\} ~=~ 1,
\end{align}
where $p_{eq}(x_0)$ is the equilibrium distribution of the system at the beginning of the protocol, $P_{\{\lambda_{m_k}\}}[x(t)|x_0]$ is the path probability (from given initial point $x_0$) for this fixed protocol, and $W$ is a function of the path.
Now we average over all possible protocols:
\begin{align}
\int &\{dm_k\}P[\{m_k\}] \int ~\mathcal{D}[x(t)]~p_{eq}(x_0)~P_{\{\lambda_{m_k}\}}[x(t)|x_0]  \nn\\
& \times \exp\left\{-\beta W[x(t), \{m_k\}]+\beta\Delta F(\lambda_{m_N}(\tau))\right\} ~=~ 1.
\label{JE1}
\end{align}
Here, $\{dm_k\}\equiv dm_1dm_2\cdots dm_{N}$, and $P[\{m_k\}]$ is the joint probability of a set of measured values $\{m_k\}$. The {\it mutual information} is defined as \cite{hor10}
\begin{equation}
I \equiv  \ln\left[\frac{p(m_1|x_1)~p(m_2|x_2)~\cdots~p(m_{N}|x_{N})}{P[\{m_k\}]}\right]. 
\label{I_mult}
\end{equation}
The path probability $P_{\{\lambda_{m_k}\}}[x(t)|x_0]$ for a fixed protocol and fixed $x_0$ is given by
\begin{align}
P_{\{\lambda_{m_k}\}}[x(t)|x_0] =& P_{\lambda_0}[x_0\to x_1] ~P_{\lambda_{m_1}}[x_1\to x_2] ~\nn\\
&\cdots ~ P_{\lambda_{m_{N}}}[x_{N}\to x_\tau],
\label{pathprob}
\end{align}
Using (\ref{I_mult}) and (\ref{pathprob}) in (\ref{JE}), we get

\begin{align}
\int \{dm_k\}\int &\mathcal{D}[x(t)] p_{eq}(x_0)~P_{\lambda_0}[x_0\to x_1]\nn\\
&~\times p(m_1|x_1)~P_{\lambda_{m_1}}[x_1\to x_2]\nn\\
&\cdots~ p(m_{N}|x_{N})~P_{\lambda_{m_{N}}}[x_{N}\to x_\tau] \nn\\
& ~\times \exp\left\{-\beta W+\beta\Delta F-I\right\} ~=~ 1.
\label{JE2}
\end{align}
To keep the notations simple, the arguments of $W$, $\Delta F$ and $I$ have been omitted. The joint probability for a phase space trajectory $x(t)$  and measured values $\{m_k\}$ in forward process (for any initial point $x_0$) is
\begin{align}
P_{\{\lambda_{m_k}\}}&[x(t),\{m_k\}] \equiv p_{eq}(x_0)~P_{\lambda_0}[x_0\to x_1] \nn\\
&\times p(m_1|x_1)~P_{\lambda_{m_1}}[x_1\to x_2]~\cdots~\nn\\
&\times p(m_{N}|x_{N})~P_{\lambda_{m_{N}}}[x_{N}\to x_\tau].
\label{Pf}
\end{align}
This is precisely the factor appearing before $e^{-\beta(W-\Delta F-I)}$ in the integrand in (\ref{JE2}). Thus we arrive at the following generalized Jarzynski Equality:
\begin{equation}
\left<e^{-\beta(W-\Delta F)-I}\right> = 1.
\label{EJE}
\end{equation}
The Jensen's inequality leads to the second law of thermodynamics which is generalized due to information gain, namely, $\la W\ra \ge \la\Delta F\ra - k_BT\la I\ra$. Since $\la I\ra\ge 0$ (being a relative entropy),  work performed on a thermodynamic system can be lowered by feedback control.

%%%%%%%%%%%%%%%%%%%%%%%%%%%%%%%%%%%%%%%%%%55555

\section{Detailed Fluctuation Theorem}
\label{ext_DFT}
The probability of forward path is given by (\ref{Pf}). To generate a reverse trajectory, we first select one of the measurement trajectories $\{m_k\}$ (with probability $P[\{m_k\}]$). We then begin with the system being at canonical equilibrium at the final value of the protocol $\lambda_{m_{N}}(\tau)$, and {\it blindly} run the full forward protocol in reverse, i.e., $\{\lambda_k(t)\} \to \{\lambda_k^\dagger(t)\} \equiv \{\lambda_k(\tau-t)\}$. We stress that no feedback is performed during the reverse process in order to respect causality \cite{hor10}. In this case, the probability of a reverse trajectory becomes
\begin{align}
P_{\{\lambda^\dagger_{m_k}\}}&[x^\dagger(t); \{m_k\}] = P[\{m_k\}]~P_{\lambda^\dagger_0}[x^\dagger_0\leftarrow x^\dagger_1]\nn\\
& P_{\lambda^\dagger_{m_1}}[x^\dagger_1\leftarrow x^\dagger_2]~\cdots~P_{\lambda^\dagger_{m_{N}}}[x^\dagger_{N}\leftarrow x^\dagger_\tau]~p_{eq}(x_\tau).
\label{Pr}
\end{align}
$P_{\{\lambda^\dagger_{m_k}\}}[x^\dagger(t); \{m_k\}]$ should not be confused with the joint probability $P_{\{\lambda^\dagger_{m_k}\}}[x^\dagger(t), \{m_k\}]$. The former represents the probability of the reverse path with fixed values $\{m_k\}$ multiplied by the probability $P[\{m_k\}]$ of choosing the set $\{m_k\}$, while the latter represents the probability of a trajectory $x(t)$ along with the measured outcomes $\{m_k\}$, if measurements are performed along the reverse process.
Now we take the ratio of (\ref{Pf}) and (\ref{Pr}):
\begin{align}
&\frac{P_{\{\lambda_{m_k}\}}[x(t),\{m_k\}]}{P_{\{\lambda^\dg_{m_k}\}}[x^\dg(t); \{m_k\}]} = \frac{P_{\{\lambda_{m_k}\}}[x(t)]}{P_{\{\lambda^\dg_{m_k}\}}[x^\dg(t)]} \nn\\
&\times \frac{p(m_1|x_1)~p(m_2|x_2)~\cdots~p(m_{N}|x_{N})}{P[\{m_k\}]} \nn\\
&= e^{\beta(W-\Delta F)} \times e^I,
\end{align}
where we have used the known Crooks' work theorem (for a predetermined protocol) \cite{cro98,cro99},
\begin{equation}
 \frac{P_{\{\lambda_{m_k}\}}[x(t)]}{P_{\{\lambda^\dg_{m_k}\}}[x^\dg(t)]} = e^{\beta(W-\Delta F)},
\end{equation}
and the definition of mutual information (eq. (\ref{I_mult})). Thus we obtain the modified detailed fluctuation theorem in presence of information,
\begin{equation}
\frac{P_{\{\lambda_{m_k}\}}[x(t),\{m_k\}]}{P_{\{\lambda^\dagger_{m_k}\}}[x^\dagger(t);\{m_k\}]} = e^{\beta(W-\Delta F) +I}.
\label{ECFT}
\end{equation}
%

%%%%%%%%%%%%%%%%%%%%%%%%%%%%%%%%%%%%

\section{Modification in Seifert's and Hatano-Sasa identities}

Now we derive other identities which are straightforward generalizations of their earlier counterparts, in the presence of information. The mathematics involved is the same as in sections \ref{ext_JE} and \ref{ext_DFT}.
For a predetermined protocol, if the pdf of the initial states for the forward path (denoted by $p_0(x_0)$) are arbitrary rather than being the Boltzmann distribution, and that of the reverse path is the final distribution of states (denoted by $p_\tau(x_\tau)$) attained in the forward path, we obtain the Seifert's theorem in lieu of the Jarzynski equality \cite{sei05,sei08}:
\begin{equation}
\la e^{-\Delta s_{tot}}\ra=1.
\label{IFT}
\end{equation}
Here, $\Delta s_{tot}=\Delta s_m+\Delta s$ is the change in the total entropy of bath and system. The path-dependent medium entropy change is given by $\Delta s_m = Q/T$, where $Q$ is the heat dissipated into the bath. $\Delta s$ is the change in the system entropy given by $\Delta s = \ln[p_0(x_0)/p_\tau(x_\tau)]$. Then eq. (\ref{IFT}) can be written as
\begin{align}
&\int ~\mathcal{D}[x(t)]~p_0(x_0)~P_{\{\lambda_{m_k}\}}[x(t)|x_0] \nn\\
&~~~~\times \exp\left\{-\Delta s_{tot}[x(t)]\right\} ~=~ 1,
\end{align}

Averaging over different sets of protocols determined by the different sets of $\{m_k\}$ values, we get
\begin{align}
\int &\{dm_k\}P[\{m_k\}] \int \mathcal{D}[x(t)]~p_0(x_0)~P_{\{\lambda_{m_k}\}}[x(t)|x_0]  \nn\\
& ~~~~~~~\times \exp\left\{-\Delta s_{tot}[x(t)]\right\} ~=~ 1.
\end{align}

Proceeding exactly in the same way as in section \ref{ext_JE} (eqns. (\ref{JE})--(\ref{EJE})), we readily get
\begin{equation}
\la e^{-\Delta s_{tot}-I}\ra=1.
\label{EIFT}
\end{equation}
Eq. (\ref{EIFT}) is the generalization of the entropy production theorem and it gives the second law in the form
\begin{equation}
\la \Delta s_{tot}\ra \ge -\la I\ra.
\end{equation}
Thus with the help of information (feedback), the lower limit of change in average entropy can be made less than zero, by an amount given by the average mutual information gained.
We can similarly show that in steady states, the detailed fluctuation theorem for total entropy becomes
\begin{equation}
\frac{P(\Delta s_{tot},I)}{P(-\Delta s_{tot},I)} = e^{\Delta s_{tot}+I}.
\label{EDFT}
\end{equation}
Here, $P(\Delta s_{tot},I)$ is the joint probability of the change in total entrop $\Delta s_{tot}$ and mutual information $I$ along the forward path.
$P(-\Delta s_{tot},I)$ is the total probability of reverse trajectories along which the change in total entropy is $-\Delta s_{tot}$, and whose corresponding forward trajectories have the mutual information $I$ between the measured and the actual states (see \cite{ran12} for details).

The Hatano-Sasa identity \cite{hat01} can also be similarly generalized:
\begin{equation}
\left<\exp\left[-\int_0^\tau dt \dot\lambda\pd{\phi(x,\lambda)}{\lambda}-I\right]\right>=1,
\label{EHSI1}
\end{equation}
where $\phi(x,\lambda) \equiv -\ln\rho_{ss}(x,\lambda)$, the negative logarithm of the nonequilibrium steady state pdf corresponding to parameter value $\lambda$. The derivations of (\ref{EDFT}) and (\ref{EHSI1}) are simple and similar to the earlier derivations (see sections \ref{ext_JE} and \ref{ext_DFT}), so they are not reproduced here. In terms of the excess heat $Q_{ex}$, which is the heat dissipated when the system moves from one steady state to another, the above equality (eq. (\ref{EHSI1})) can be rewritten in the following form (for details see \cite{hat01}):
\begin{equation}
\la\exp[-\beta Q_{ex}-\Delta\phi-I]\ra=1.
\end{equation}
Using the Jensen's inequality, the generalized second law for transitions between nonequilibrium steady states follows, namely, 
\begin{equation}
T\la\Delta s\ra \ge -\la Q_{ex}\ra  -k_BT\la I\ra,
\end{equation}
where $\Delta s$ is the change in system entropy defined by $\Delta s\equiv -\ln\frac{\rho_{ss}(x,\lambda_\tau)}{\rho_{ss}(x,\lambda_0)} = \Delta \phi$.

%%%%%%%%%%%%%%%%%%%%%%%%%%%%%%%%555
\section{The generalized Jarzynski equality and the efficacy parameter}
\label{CEP}
The Jarzynski equality can also be extended to a different form in the presence of information:
\begin{equation}
\la e^{-\beta(W-\Delta F)}\ra = \gamma,
\end{equation}
where $\gamma$ is the efficacy parameter \cite{sag10,sag11}. 
The efficacy parameter $\gamma$ defines how efficient our feedback control is. 
Following similar mathematical treatment as  in the derivation of extended Jarzynski equality, we have
\begin{align}
\gamma=& \int \mathcal{D}[x(t)]\{dm_k\} P_{\{\lambda_{m_k}\}}[x(t),\{m_k\}] e^{-\beta(W-\Delta F)} \nn\\
=& \int \mathcal{D}[x(t)]\{dm_k\} P_{\{\lambda^\dagger_{m_k}\}}[x^\dagger(t);\{m_k\}] ~e^I .
\end{align}
In the last step we have used relation (\ref{ECFT}). Using  the definition (\ref{Pr}) for the reverse trajectory and eq. (\ref{I_mult}) for the mutual information, we get
\begin{align}
\gamma &= \int \mathcal{D}[x(t)]\{dm_k\} P[\{m_k\}]~P_{\lambda^\dagger_0}[x^\dagger_0\leftarrow x^\dagger_1]\nn\\
& P_{\lambda^\dagger_{m_1}}[x^\dagger_1\leftarrow x^\dagger_2]~\cdots~P_{\lambda^\dagger_{m_{N}}}[x^\dagger_{N}\leftarrow x^\dagger_\tau]~p_{eq}(x_\tau) \nn\\
& \times \frac{p(m_1|x_1)~p(m_2|x_2)~\cdots~p(m_{N}|x_{N})}{P[\{m_k\}]},
\end{align}
Now we use the assumption of  time-reversal symmetry of the measurements \cite{sag10}, i.e., $p(m_k^\dg|x_k^\dg)=p(m_k|x_k)$.  This leads to
\begin{align}
\gamma &= \int \mathcal{D}[x(t)]\{dm_k\} P_{\{\lambda^\dagger_{m_k}\}}[x^\dagger(t)] \nn\\
&~~\times p(m_1^\dagger|x_1^\dagger)~p(m_2^\dagger|x_2^\dagger) ~\cdots~p(m^\dagger_{N}|x^\dagger_{\tau}) \nn\\
&= \int \mathcal{D}[x^\dagger(t)]~\{dm_k\}~ P_{\{\lambda^\dagger_{m_k}\}}[x^\dagger(t), \{m^\dagger_k\}].
\end{align}
Here, $P_{\{\lambda^\dagger_{m_k}\}}[x^\dagger(t), \{m^\dagger_k\}]$ is the joint probability of obtaining both the trajectory $x(t)$ and the set of measured outcomes $\{m^\dg_k\}$ in the reverse process under the protocol $\{\lambda^\dg_{m_k}\}$. Finally, performing the integral over the trajectories, we arrive at
\begin{align}
\gamma = \int \{dm_k\} ~P_{\{\lambda^\dagger_{m_k}\}}[\{m_k^\dagger\}].
\end{align}
Here, $P_{\{\lambda^\dagger_{m_k}\}}[\{m^\dagger_k\}]$ is the probability of observing the measurement trajectory $\{m_k^\dg\}$ for the reverse process.
Physically, this means that $\gamma$ also describes the sum of the  probabilities of observing  {\it time-reversed} outcomes in the time-reverse protocols, for all set of possible protocols.

%%%%%%%%%%%%%%%%%%%%%%%%%%%%%%%%%%%%%%%55
\section{Quantum case}

Now we extend the above treatment to the case of a quantum open system. Hanggi et al have shown \cite{han10} that for a closed quantum system, the fluctuation theorems remain unaffected even if projective measurements are performed in-between. This happens in spite of the fact that the probabilities of the forward and backward paths (by ``path'' we mean here a collection of the successive eigenstates to which the system collapses each time a projective measurement is performed) do change in general. Taking cue from this result, we proceed as follows.
The supersystem consisting of the bath and the system evolve under the total Hamiltonian
\begin{equation}
H(t) = H_S(t) + H_{SB} + H_B.
\end{equation}
The bath Hamiltonian $H_B$ and the interaction Hamiltonian $H_{SB}$ have been assumed to be time-independent, whereas the system Hamiltonian $H_S(t)$ depends explicitly on time through a time-dependent external drive $\lambda(t)$.
We first prepare the supersystem at canonical equilibrium  at temperature $T$. At initial time  $t_0=0$, we measure the state of the total system and the collapsed eigenstate is $|i_0\ra$. The notation means that the total system has collapsed to the $i_0^{th}$ eigenstate (of the corresponding measured operator, which is the total Hamiltonian $H(0)$ at $t=0$) when measured at time $t=0$. 
  The supersystem consisting of the bath and the system is described by
\begin{equation}
\rho(0) \equiv \frac{e^{-\beta H(0)}}{Y(0)} ~~\Ra ~~ \rho_{i_0i_0} = \frac{e^{-\beta E_{i_0}}}{Y(0)}
\end{equation}
In the above relation, $\rho_{i_0i_0}$ are the diagonal elements of the initial density matrix of the supersystem, and $Y(0)$ is the partition function for the entire supersystem:
\begin{equation}
Y(0) = \Tr{} e^{-\beta H(0)}.
\end{equation}

We then evolve the system up to time $t_1$ under a predetermined protocol $\lambda_0(t)$, and at $t_1$ we once again measure some observable $M$ of the system. Let the outcome be $m_1$, whereas the actual collapsed state is $|i_1\ra$ corresponding to eigenvalue $M_{i_1}$. This outcome is obtained with probability $p(m_1|i_1)$, which is an assumed classical error involved in the measurement. Now we introduce the feedback by modifying the original protocol to $\lambda_{m_1}(t)$, and then continue up to $t_2$, where we perform the measurement to get outcome $m_2$, the actual state being $|i_2\ra$. Subsequently our protocol becomes $\lambda_{m_2}(t)$, and so on. Thus the probability of getting the set of eigenstates $\{i_k\}\equiv \{|i_0\ra, |i_1\ra, \cdots, |i_\tau\ra\}$ with the measurement trajectory $\{m_k\}$ is given by %
\begin{align}
P_{\{\lambda_{m_k}\}}&[\{i_k\},\{m_k\}] = \rho_{i_0i_0}~|\la i_1|U_{\lambda_0}(t_1,0)|i_0\ra|^2 \nn\\
&\times p(m_1|i_1)~|\la i_2|U_{\lambda_{m_1}}(t_2,t_1)|i_1\ra|^2 \times \cdots \nn\\
&\times p(m_{N}|i_{N})~|\la i_\tau|U_{\lambda_{m_{N}}}(\tau,t_{N})|i_N\ra|^2.
\label{Pf_q}
\end{align}
Here, $U_{\lambda_{m_i}}(t_{i+1},t_i)$ is the unitary evolution operator from time $t_i$ to time $t_{i+1}$.
The reverse process is generated by starting with the supersystem in canonical equilibrium with protocol value $\lambda_{m_N}(\tau)$, and blindly reversing one of the chosen protocols of the forward process. Now we need to perform measurements along the reverse process as well, simply to ensure that the state {\it does} collapse to specific eigenstates and we do obtain an unambiguous reverse trajectory in each experiment.  However, in order to respect causality \cite{hor10}, we do not perform feedback during the reverse process. The probability for a trajectory that starts from initial collapsed energy eigenstate $|i_\tau\ra$  and follows the exact sequence of collapsed states as the forward process is given by
\begin{align}
P_{\{\lambda^\dg_{m_k}\}}&[\{i_k\};\{m_k\}] = |\la i_0|\Theta^\dg U_{\lambda^\dg_0}(\tilde 0,\tilde t_1)\Theta|i_1\ra|^2 \nn\\
&\times |\la i_1|\Theta^\dg U_{\lambda^\dg_{m_1}}(\tilde t_1,\tilde t_2)\Theta|i_2\ra|^2 \times \cdots \nn\\
&\times |\la i_N|\Theta^\dg U_{\lambda^\dg_0}(\tilde t_N,\tilde \tau)\Theta|i_\tau\ra|^2~\rho_{i_\tau i_\tau} ~P[\{m_k\}].
\label{Pr_q}
\end{align}
Here, $\Theta$ is a time-reversal operator \cite{han10}, and $|\la i_k|\Theta^\dg U_{\lambda_{m_k}^\dg}(\tilde t_k,\tilde t_{k+1})\Theta|i_{k+1}\ra|^2$ is the probability of transition from the time-reversed state $\Theta |i_{k+1}\ra$ to $\Theta |i_{k}\ra$ under the unitary evolution with the reverse protocol: $U_{\lambda_{m_k}}(\tilde t_k,\tilde t_{k+1}) = U_{\lambda_{m_k}}(\tau-t_k,\tau-t_{k+1})$.
Here, the tilde symbol implies time calculated along the reverse trajectory: $\tilde t \equiv \tau-t$. $\rho_{i_\tau i_\tau}$ is the diagonal element of the density matrix when the system is at canonical equilibrium at the beginning of the reverse process:
\begin{equation}
\rho_{i_\tau i_\tau} = \frac{e^{-\beta E_{i_\tau}}}{Y(\tau)}.
\end{equation}
Now we have,
\begin{align}
&\Theta^\dagger U_{\lambda_{m_k}^\dagger}(\tilde t_k,\tilde t_{k+1})\Theta \nn\\
&= \Theta^\dg T \exp\left[-\frac{i}{\hbar}\int_{\tilde t_{k+1}}^{\tilde t_k}H_{\lambda_{m_k}^\dg}(t)dt\right]\Theta \nn\\
&= T\exp\left[+\frac{i}{\hbar}\int_{\tilde t_{k+1}}^{\tilde t_k}H_{\lambda_{m_k}^\dg}(t)dt\right],
\end{align}
where $T$ implies time-ordering.
Changing the variable $t\to \tau-t$, we get
\begin{align}
&\Theta^\dagger U_{\lambda_{m_k}^\dagger}(\tilde t_k,\tilde t_{k+1})\Theta \nn\\
& = T\exp\left[-\frac{i}{\hbar}\int_{ t_{k+1}}^{ t_k}H_{\lambda_{m_k}^\dg}(\tau-t)dt\right] \nn\\
& = T\exp\left[-\frac{i}{\hbar}\int_{ t_{k+1}}^{ t_k}H_{\lambda_{m_k}}(t)dt\right] \nn\\
&= U_{\lambda_{m_k}}(t_k,t_{k+1}) = U^\dg_{\lambda_{m_k}}(t_{k+1},t_k).
\end{align}

Accordingly, 
\begin{align}
&\la i_k |\Theta^\dg U_{\lambda_{m_k}^\dg}(\tilde t_k, \tilde t_{k+1})\Theta |i_{k+1}\ra \nn\\
=& \la i_k|U^\dg_{\lambda_{m_k}}(t_{k+1},t_k)|i_{k+1}\ra = \la i_{k+1}| U_{\lambda_{m_k}}(t_{k+1},t_k)|i_k\ra^\dg.
\end{align}

Thus, while dividing (\ref{Pf_q}) by (\ref{Pr_q}), all the modulus squared terms cancel, and we have,
\begin{align}
\frac{P_{\{\lambda_{m_k}\}}[\{i_k\},\{m_k\}]}{P_{\{\lambda^\dg_{m_k}\}}[\{i_k\};\{m_k\}]} &= \frac{\rho_{i_0 i_0}}{\rho_{i_\tau i_\tau}}~ \frac{p(m_1|i_1)\cdots p(m_N|i_N)}{P[\{m_k\}]} \nn\\
&= \frac{Y(\tau)}{Y(0)}~e^{\beta W+I},
\label{DFT_q}
\end{align}
where $W\equiv E_{i_\tau}-E_{i_0}$ is the work done by the external drive on the system. 
This follows from the fact that the external forces act only on the system $S$.
Now we follow \cite{han09} and define the equilibrium free energy of the system, $F_S(t)$, as the thermodynamic free energy of the open system, which is the difference between the total free energy $F(t)$ of the supersystem and the bare bath free energy $F_B$:
\begin{equation}
F_S(t) \equiv F(t) - F_B.
\end{equation}
Here, $t$ specifies the values of the external parameters in the course of the protocol at time $t$. From the above equation, the partition function for the open system is given by \cite{han09}
\begin{equation}
Z_s(t) = \frac{Y(t)}{Z_B} = \frac{\Tr{S,B}e^{-\beta H(t)}}{\Tr{B}e^{-\beta H_B}},
\label{Z_S}
\end{equation}
where $S$ and $B$ represent system and bath variables, respectively.
Since $Z_B$ is independent of time, using (\ref{Z_S}) in (\ref{DFT_q}), we have
\begin{equation}
\frac{P_{\{\lambda_{m_k}\}}[\{i_k\},\{m_k\}]}{P_{\{\lambda^\dg_{m_k}\}}[\{i_k\};\{m_k\}]} = \frac{Z_S(\tau)}{Z_S(0)}~e^{\beta W+I} = e^{\beta(W-\Delta F_S)+I}.
\label{EQFT}
\end{equation}

The above relation is the extended form of the Tasaki-Crooks detailed fluctuation theorem for open quantum systems in presence of feedback where the measurement process involves classical errors.

From (\ref{EQFT}), the quantum Jarzynski Equality follows:
\begin{align}
&\sum_{\{i_k\}}\int \{dm_k\} P_{\{\lambda_{m_k}\}}[\{i_k\},\{m_k\}] e^{-\beta(W-\Delta F_S)-I} \nn\\
&= \sum_{\{i_k\}}\int \{dm_k\} P_{\{\lambda^\dg_{m_k}\}}[\{i_k\};\{m_k\}] = 1, \nn
\end{align}
i.e.,
\begin{equation}
\left<e^{-\beta(W-\Delta F_S)-I}\right> = 1.
\end{equation}
This is valid for open quantum system and is independent of the coupling strength and the nature of the bath.

The quantum efficacy parameter is defined as $\la e^{-\beta(W-\Delta F_S)}\ra \equiv \gamma$, and the calculation of $\gamma$ is exactly in the spirit of section \ref{CEP}, except that $\int\mathcal{D}[x(t)]$ is replaced by $\sum_{\{i_k\}}$, i.e., summation over all possible eigenstates. Finally we get the same result, namely, 
\begin{equation}
\gamma = \int \{dm_k\}P_{\{\lambda^\dg_{m_k}\}}[\{m^\dg_k\}].
\end{equation}
%

%%%%%%%%%%%%%%%%%%%%%%%%%%%%%%%%

\section{Discussion and conclusions}

In conclusion, we have rederived several extended fluctuation theorems in the presence of feedback. To this end, we have used  several equalities given by the already known fluctuation theorems. We have extended the quantum fluctuation theorems for open systems, following the earlier treatment \cite{han10,han09}. No assumption is made on the strength of the system-bath coupling and the nature of the environment. However, we have assumed that the measurement process leading to information gain involves classical errors.
The fluctuation-dissipation theorem is violated in nonequilibrium systems. The effects of feedback on this violation are being studied \cite{ito11}. This is the subject of our ongoing work.

\section{Acknowledgement}

One of us (AMJ) thanks DST, India for financial support.

%%%%%%%%%%%%%%%%%%%%%%%%%%%%%%%%%%%%5

\end{document}